\documentclass[twocolumn]{article}
\usepackage{graphicx}

\newcommand{\eq}[1]{Eq.~(\ref{eq.#1})} 

\newcommand{\fig}[1]{Fig.~\ref{fig.#1}}
\newcommand{\figbare}[1]{\ref{fig.#1}}

\newcommand{\tbl}[1]{Table~\ref{table.#1}}

\newcommand{\sect}[1]{Sec.~\ref{sect.#1}}

\newcommand{\sectlabel}[1]{\label{sect.#1}}
\newcommand{\eqlabel}[1]{\label{eq.#1}}
\newcommand{\figlabel}[1]{\label{fig.#1}}
\newcommand{\tbllabel}[1]{\label{table.#1}}

\newcommand{\figwidth}{3in}

\newcommand{\Puser}{P_{\rm user}}
\newcommand{\Presolve}{P_{\rm resolve}}
\newcommand{\mean}[1]{{\hat{#1}}}
\newcommand{\vAge}{v^{\rm age}} 
\newcommand{\vResolve}{v^{\rm resolve}} 

\begin{document}

\title{Diversity of Online Community Activities}
\author{Tad Hogg\\HP Labs\\Palo Alto, CA 94304 \and
Gabor Szabo\\HP Labs\\Palo Alto, CA 94304}

\maketitle
\begin{abstract}
Web sites where users create and rate content as well as form networks
with other users display long-tailed distributions in many aspects of
behavior. Using behavior on one such community site, Essembly, we
propose and evaluate plausible mechanisms to explain these behaviors.
Unlike purely descriptive models, these mechanisms rely on user
behaviors based on information available locally to each user. For
Essembly, we find the long-tails arise from large differences among
user activity rates and qualities of the rated content, as well as the
extensive variability in the time users devote to the site. We show
that the models not only explain overall behavior but also allow
estimating the quality of content from their early behaviors.
\end{abstract}

\section{Introduction}

Participatory web sites facilitate their users creating, rating and
sharing content. Examples include Digg[.com] for news stories,
Flickr[.com] for photos and Wikipedia[.org] for encyclopedia
articles. To aid users in finding content, many such sites employ
collaborative filtering~\cite{lam04} to allow users to specify links
to other users whose content or ratings are particularly relevant.
These links can involve either people who already know each other
(e.g., friends) or people who discover their common interests
through participating in the web site. In addition to helping
identify relevant content, the resulting networks enable users to
find others with similar interests and establish trust in
recommendations~\cite{guha04}.

The availability of activity records from these sites has led to
numerous studies of user behavior and the networks they create.
Observed commonalities in these systems suggest general generative
processes leading to these observations. Examples include
preferential attachment in forming networks and multiplicative
processes leading to wide variation in user activity. While such
models provide a broad understanding of the observations, they often
lack causal connection with plausible user behaviors based on user
preferences and the information available to users in making their
decisions~\cite{vazquez03,boccaletti06}. Moreover, observed behavior
can arise from a variety of mechanisms~\cite{mitzenmacher04}.

For predicting consequences of alternate designs of the web site,
models including causal behavior are necessary. Establishing such
models is more difficult than simply observing behavior: due to the
possibility of confounding factors in observations, many different
causal models can produce the same observations. Instead, such
models would ideally use intervention studies and randomized trials
to identify important causal relationships. In contrast to the wide
availability of observational data on user behavior, such
intervention studies are difficult, though this is situation is
improving with the increasing feasibility of experiments in large
virtual communities~\cite{bainbridge07} and large-scale web-based
experiments~\cite{salganik06}.

Nevertheless, identifying information readily available to users on
a participatory web site can suggest plausible causal mechanisms.
Such models provide specific hypotheses to test with future
intervention experiments and also suggest improvements to overall
system behavior by altering the user experience, e.g., available
information or incentives. The simplest such approach considers
average behavior of users on a site~\cite{lerman07}. Such models can
indicate how system behavior relates to the average decisions of
many users. By design, such models do not address a prominent aspect
of observed online networks: the long tails in their distributions
of links and activity. Models including this diversity could be
useful to improve effectiveness of the web sites by allowing focus
on significantly active users or especially interesting content, and
enhancing user experience by leveraging the long tail in niche
demand~\cite{anderson06}.

A key question with respect to the observed diversity is whether
users, content and the networks are reasonably viewed as behaviors
arising from a statistically homogeneous population, and hence
well-characterized by a mean and variance. Or is diversity of
intrinsic characteristics among participants the dominant cause of
the observed wide variation in behaviors? In the latter case, can
these characteristics be estimated (quickly) from (a few)
observations of behavior, allowing site design to use estimates of
these characteristics, e.g., to highlight especially interesting
content? Moreover, to the extent user diversity is important, what
is a minimal characterization of this user variation sufficient to
produce the observed long-tail distributions?

This paper considers these questions in the context of a
politically-oriented web community, Essembly\footnote{Essembly LLC at
  www.essembly.com}. Unlike most such sites, Essembly provides
multiple networks with differing nominal semantics, which is useful
for distinguishing among some models. We consider plausible mechanisms
users could be following to produce the observed long-tail behaviors
both in their online activities and network characteristics. In the
remainder of this paper, we first describe Essembly and our data set
in \sect{essembly}. We then separately examine highly variable
behaviors for users, content rating and network formation, in
Secs.~\ref{sect.users}, \ref{sect.resolves}, and \ref{sect.links},
respectively. We suggest models to describe the observed
characteristics of users, content, and the network, and consider their
possible use during operation of the web site by helping identify user
and content parameters early in their history. Finally we discuss
implications and extensions to other participatory web sites.

In the three sections focusing on user behavior, resolve
characteristics, and network structure, respectively, we first
introduce the observations, then present a model describing these
observations (subsections \emph{Model}), and finally analyze the model
parameters and predictions (subsections \emph{Behavior}).

\section{Essembly}
\sectlabel{essembly}

Essembly is an online service helping users engage in political
discussion through creating and voting on \emph{resolves} reflecting
controversial issues. Essembly provides three distinct networks for
users: a social network, an ideological preference network, and an
anti-preference network, called \emph{friends} (those who know each
other in person), \emph{allies} (those who share similar ideologies)
and \emph{nemeses} (those who have opposing ideologies), respectively.

The distinct social and ideological networks enable users to
distinguish between people they know personally and people encountered
on the site with whom they tend to agree or disagree. Network links
are formed by invitation only and each link must be approved by the
invitee. Thus all three networks in Essembly are explicitly created by
users. Essembly provides a ranked list of ideologically most similar
or dissimilar users based on voting history, thus users can identify
potential allies or nemeses by comparing profiles. With regards to
voting activity, the Essembly user interface presents several options
for users to discover new resolves, for instance based on votes by
network neighbors, recency, overall popularity, and degree of
controversy.

Our data set consists of anonymized voting records for Essembly
between its inception in August 2005 and December 2006, and the users
and the links they have at that time in the three networks at the end
of this period. Our data set has $15,424$ users. Essembly presents 10
resolves during the user registration process to establish an initial
ideological profile used to facilitate users finding others with
similar or different political views. To focus on user-created
content, we consider the remaining $24,953$ resolves, with a total of
$1.3$ million votes.

\section{Users}\sectlabel{users}

\begin{figure}[t]
\centering
\includegraphics[width=\figwidth]{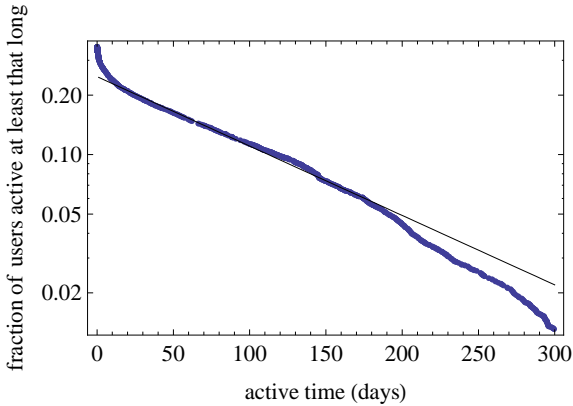}
\caption{\figlabel{activity time}Distribution of activity times for
users. The line shows an exponential fit to the values between 10
and 200 days, proportional to $e^{-t/\tau}$ where $\tau=124\mbox{
days}$.}
\end{figure}

\begin{figure}[t]
\centering
\includegraphics[width=\figwidth]{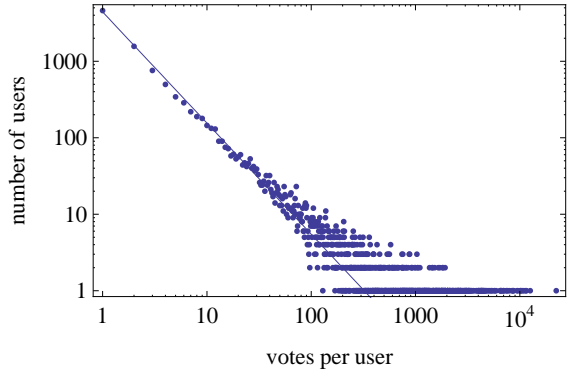}
\caption{\figlabel{votes per user}Distribution of number of users
vs.~the number of votes a user made. The solid curve indicates a
Zipf distribution fit to the values, with parameter $\nu=0.45\pm
0.01$. In this and other figures the range given with the parameter
estimate is the $95\%$ confidence interval. The plot does not
include the $2984$ users with zero votes.}
\end{figure}

\fig{activity time} shows most users are active for only a short
time (less than a day), as measured by the time between their first
and last votes (this includes votes on the initial resolves during
registration -- users need not vote on all of them immediately). The
4762 users active for at least a day account for most of the votes
and links, and we focus on these \emph{active users} for our model.
For these users, \fig{activity time} shows an exponential fit to the
activity distribution for intermediate times. Thus users who have
sufficient interest in the system to participate for at least a few
days behave approximately as if they decide to stop participating as
a Poisson process. The additional decrease at long times (above 200
days or so) is due to the finite length of our data sample (about
500 days).

About a fifth of users have no votes on noninitial resolves. For the
rest of the users, \fig{votes per user} shows the distribution of
votes among users who voted at least once for noninitial resolves.
These votes are close to a Zipf distribution in number of votes, with
number of users with $v$ votes proportional to $v^{-\nu-1}$. The
parameter estimates and confidence intervals in this and the other
figures are maximum likelihood estimates~\cite{newman05,james07}
assuming independent samples. This wide variation in user activity
also occurs in other participatory web sites such as
Digg~\cite{lerman07a}.

The distribution of number votes per user arises from two factors:
how long users participate before becoming inactive, and how often
they vote while active.

\subsection{Model}

\begin{figure}[t]
\centering \includegraphics[width=3in]{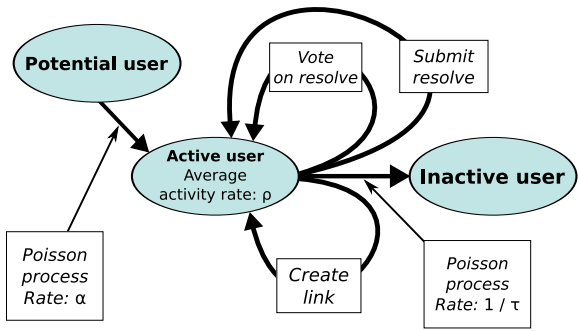}
\caption{\figlabel{user model}Model of user behavior. People join
the site as active users, who \emph{create} resolves, \emph{vote} on
them and \emph{link} to other active users. Users can eventually
stop participating and become inactive.}
\end{figure}

\fig{user model} summarizes our model for user behavior. This models
the participation of users and their activities on the site while
they are active. New users arrive in the system when they register,
and we model this as a Poisson process with rate $\alpha$, and such
users leave the system (\emph{i.e.}, become inactive) with a rate
$1/\tau$. \tbl{user parameters} gives the values for these model
parameters based on average arrival and activity times of active
users.

\begin{table}[t]
\centering
\begin{tabular}{ll}
parameter & value \\ \hline
new user rate & $\alpha = 9.3/\mbox{day}$ \\
activity time constant  & $\tau = 124\mbox{ days}$ \\
resolve creation    & $q=0.018\pm 0.0002$ \\
link creation       & $\lambda=0.043\pm 0.0003$
\end{tabular}
\caption{\tbllabel{user parameters}User activity parameters.}
\end{table}

User activities consist of \emph{voting}, \emph{creating resolves},
and \emph{forming links}. User activity is clumped in time, with
groups of many votes close in time separated by gaps of at least
several hours. This temporal structure can be viewed as a sequence
of user sessions. The averaged distributions for interevent times
between activities of individuals show long-tail behavior, similarly
to other observed human activity patterns, such as email
communications or web site visits~\cite{vazquez06}. To model the
number of votes per user in the long time limit where we are only
interested in the total number of accumulated votes for a particular
user, this clumping of votes in time is not important. Specifically
we suppose each user has an average activity rate $\rho$ while they
are active on the site (cf. \fig{activity time}), given as $\rho_u =
e_u / T_u$, where $\rho_u$ is user $u$'s activity, $e_u$ is her
number of events (i.e., votes, resolve creations and links), and
$T_u$ is the time elapsed between her first and last vote. We
suppose the $\rho_u$ values arise as independent choices from a
distribution $\Puser(\rho_u)$ and the values are independent of the
length of time a user is active on the site. These properties are
only weakly correlated (correlation coefficient $-0.06$ among active
users).

We characterize user activities by fractions $q$ and $\lambda$ for
creating resolves and forming links, respectively. The rate of
voting on existing resolves for a user is then
$\rho_u(1-q-\lambda)$, which is by far the most common of the three
user activities. For simplicity, we treat these choices as
independent and take $q$ and $\lambda$ to be the same for all users.
Thus in our model, the variation among users is due to their
differing overall activity rates $\rho_u$ and amount of time they
are active on the site $T_u$.

\subsection{Behavior}

\begin{figure}[t]
\centering \includegraphics[width=\figwidth]{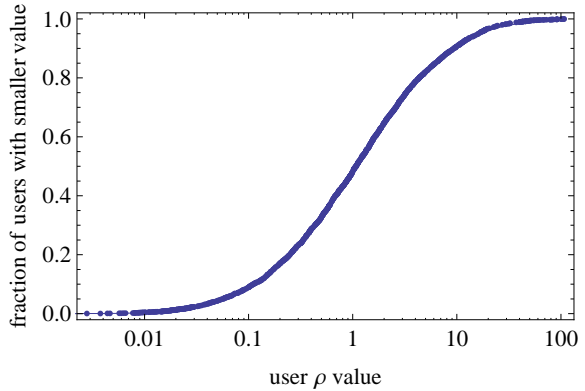}
\caption{\figlabel{rho estimates}Cumulative distribution of activity
rates, $\rho_u$, for the 4719 users who were active at least one day
and voted
  on at least one noninitial resolve or formed at least on link. Plot includes a curve for a
  lognormal distribution fit, which is indistinguishable from the
  points and with parameters $\mu=0.03\pm 0.05$ and $\sigma=1.70\pm 0.04$. The
  $\rho$ values are in units of actions per day.}
\end{figure}

We estimate the model parameters from the observed user activities,
and restrict attention to active users. \tbl{user parameters} shows
the estimates for parameters, $q$ and $\lambda$, governing activity
choices. \fig{rho estimates} shows the observed cumulative
distribution $\rho_u$ values and a fit to a lognormal distribution.

The heavy tailed nature of the votes per user distribution
(\fig{votes per user}) can be attributed to the interplay between
the user activity times $T_u$ and the broad lognormal distribution
of the user activity rates $\rho_u$: the mixture of these two
distributions results in a power law, as has been shown in the
context of web page links as well~\cite{huberman98}.

The distributions of activity times and rates presumably reflect the
range of dedication of users to the site, where most users are
trying the service for a very limited time but active users are also
represented in the heavy tail. Such extended distributions of user
activity rates is also seen in other activities, including use of
web sites, e.g., Digg~\cite{lerman07a}, and scientific
productivity~\cite{shockley57}.

\section{Resolves}\sectlabel{resolves}

A key question for user-created content is how user activities
distribute among the available content. For Essembly, \fig{votes per
  resolve} shows the total number of votes per resolve. This
distribution covers a wide range, with some resolves receiving many
times as many votes as the median. In Essembly, each resolve receives
its first vote when it is created, i.e., the vote of the user
introducing the resolve. Thus the observed votes on a resolve are a
combination of two user activities: creating a new resolve (giving the
resolve its first vote) and subsequently other users choosing to vote
on the resolve if they see it while visiting the site. We note that
users do not see the distribution of previous votes until they cast
their votes, so that their judgement is unbiased. After voting, they
can see how other users had voted on the resolve.

\begin{figure}[t]
\centering
\includegraphics[width=\figwidth]{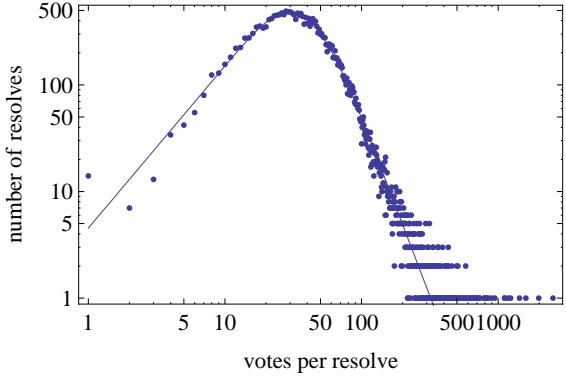}
\caption{\figlabel{votes per resolve}Distribution of votes on
resolves. The solid curve indicates a double Pareto lognormal fit to
the values, with parameters $\alpha=2.4\pm 0.1$, $\beta=2.5 \pm
0.1$, $\mu=3.67\pm 0.02$ and $\sigma=0.38 \pm 0.02$.}
\end{figure}

We consider a user's selection of an existing resolve to vote on as
mainly due to a combination of two factors: visibility and
interestingness of a resolve to a user. Visibility is the
probability a user finds the resolve during a visit to the site.
Interestingness is the conditional probability a user votes on the
resolve given it is visible to that user. These two factors apply to
a variety of web sites, e.g., providing a description of average
behavior on Digg~\cite{lerman07}.

The design of the web site's user interface determines content
visibility. Typically sites, including Essembly, emphasize recently
created content and popular content (i.e., receiving many votes over
a period of time). Essembly also emphasizes controversial resolves.
As with other networking sites, the user interface highlights
resolves with these properties both globally and among the user's
network neighbors. Users can also find resolves through a search
interface. While we cannot observe which resolves people click on, we
do register when they vote on them, and thus find it interesting
enough to warrant spending time to consider them.

In a similar vein, clickthrough rates have been extensively
investigated in the context of web search and search engine
optimization. Web search engines strive to provide users with relevant
results to their queries, and rank the matching documents in reverse
order of perceived importance to the searcher. However, due to the
fact that search queries are not well defined and several possible
optimal results may match a user's request, it is not always the top
ranked result that is most relevant to the user. Search engine logs
provide data on which results users click on for given queries, and
thus can reveal users' implicit relevance judgements to their
searches. It has been shown, however, that the probability of clicking
on a given result is biased by the presentation order, thus a result
with the same relevance as another but appearing in a higher poisition
may get more clicks (this is also called ``trust
bias'')~\cite{craswell08,clarke07}. Eyetracking experiments have also
shown that users scan through search results in a linear order from
top to bottom, which further explains why results on the top are
clicked with a larger probability~\cite{joachims05}. Clickthroughs are
analogous to votes cast on resolves in Essembly, indicating a
preference on the part of the user for the given item found for the
query during a web search, and the resolve voted on in Essembly,
respectively. Predictive models have been developed to compensate for
position bias and to offset it to reveal the true relevance of the
search results for the users~\cite{agichtein06,carterette07}.

In Essembly, recency appears to be the most significant factor
affecting visibility, in a very similar manner to how search engine
users perceive the ranked results. \fig{votes vs age} shows how votes
distribute according to the age of the resolve at the time of the
vote. We define the \emph{age} of the resolve as the ordinality of the
given resolve among resolves introduced in time. An age 1 resolve is
the newest one of the resolves introduced, while the oldest resolve
has age $R$ where $R$ is the number of resolves. Most votes go to
recent resolves with a small age.

\begin{figure}[t]
\centering
\includegraphics[width=\figwidth]{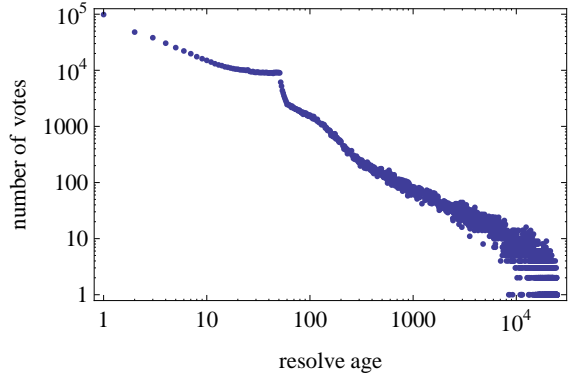}
\caption{\figlabel{votes vs age}Distribution of votes vs. age of a
resolve.}
\end{figure}

The decay in votes with age is motivated by recency (decreasing
visibility with age as resolve moves down, and eventually off, the
list of recent resolves). We offer no underlying model for this
``aging function'' but its overall power-law form corresponds to
users' willingness to visit successive pages or scroll down a long
list~\cite{huberman98}. The step at age 50 is, presumably, due to a
limit on number of recent resolves readily accessible to users. The
values decrease as a power law, proportional to $a^{-s}$, where $a$ is
resolve age and $s\approx 0.5$ up to about age 50. For larger ages,
the values in \fig{votes vs age} decreases faster, with $s\approx-1$.
It has also been found that in search engine result pages the
probability of clicking on a result also decreases with the rank of
the result as a power law, albeit with a different exponent
($-1.6$)~\cite{lempel03,fortunato06}.

The combination of different ages in the data sample is a significant
factor in producing the observed distributions~\cite{huberman99a}. In
particular, a distribution of ages and a multiplicative process
produces a lognormal distribution with power-law tails, the double
Pareto lognormal distribution~\cite{reed04}, with four parameters. Two
parameters, $\mu$ and $\sigma$ characterize the location and width of
the center of the distribution. The remaining parameters characterize
the tails: $\alpha$ for the power-law decay in the upper tail, with
number of resolves with $v$ votes proportional to $v^{-\alpha-1}$, and
$\beta$ for the power-law growth in the lower tail, with number of
resolves proportional to $v^{\beta-1}$. \fig{votes per resolve} shows
a fit of this distribution to the numbers of votes different resolves
received. For Essembly, the networks have only a modest influence on
voting~\cite{hogg08a}.

\subsection{Model}

Our model of resolve creation, described in \sect{users}, involves a
fraction $q$ of each user's activity on the site, on average, giving
each resolve its first vote. For subsequent votes, we view a user's
choice of resolve as due to an intrinsic \emph{interestingness}
property $r$ of each resolve and its visibility.

In general $r$ could depend on the resolve age and its popularity
(especially among network neighbors, if neighbors influence a user to
vote rather than just make a resolve more visible). However, for
simplicity, we take $r$ to be constant for a resolve. A key motivation
for this choice is the observation that high or low rates of voting on
a resolve tend to persist over time, when controlling for the age and
number of votes the resolve already has. Thus the importance of an
intrinsic interestingness property of resolves is a reasonable
approximation for Essembly (as discussed further in \sect{online
  estimation}). We further assume $r$ is independent of the user,
which amounts to considering general interest in resolves among the
population rather than considering possible niche interests among
subgroups of users. With these simplifications, we take the $r$ values
to arise as independent choices from a distribution $\Presolve(r)$.

Visibility of a resolve depends on age, rank in number of votes
compared with other resolves (popularity), controversy, both in
general and among user's neighbors. For Essembly, resolve age appears
to be the most significant factor, so we take visibility to be a
function of age $a$ alone, as determined by a function $f(a)$.

With these factors, we model the chance that the next vote on
existing resolves goes to resolve $j$ as being proportional to $r_j
f(a_j)$ where $a_j$ is the age of the resolve at the time of the
vote. The model's behavior is unchanged by an overall multiplicative
constant, and we arbitrarily set $f(1)=1$.

\subsection{Behavior}\sectlabel{resolves behavior}

We would like to estimate the distribution $\Presolve$ and the aging
function $f(a)$. To do so, we consider the votes (other than the first
vote on each resolve) between successive resolve introductions.
Specifically, let $R$ be the number of resolves in our data sample. We
denote the resolves in the order they were introduced, ranging from 1
to $R$.

Let us assume that there have $i$ resolves been introduced in Essembly
up to a given time, and let $v_i$ be the number of votes made in the
time interval $I_i$ between the introductions of resolves $i$ and
$i+1$ (not including the two votes accompanying those resolve
introductions). During this interval, the system has $i$ existing
resolves as assumed. When the number of existing resolves is large, we
can treat the votes going to each resolve as approximately
independent. In this case, the number of votes resolve $j\leq i$
receives during time interval $I_i$ is a Poisson process with mean
$v_i r_j f(i-j+1)$ because during this interval resolve $j$ is of age
$i-j+1$. \tbl{r estimate} illustrates these relationships.

\begin{table}[t]
\centering
\begin{displaymath}
\begin{array}{c|cccc}
            & \multicolumn{4}{c}{\mbox{interval}} \\
 \mbox{resolve} & I_1 & I_2 & I_3 & I_4 \\ \hline
 1 & f(1) r_1 v_1 & f(2) r_1 v_2 & f(3) r_1 v_3 & f(4) r_1 v_4 \\
 2 & - & f(1) r_2 v_2 & f(2) r_2 v_3 & f(3) r_2 v_4 \\
 3 & - & - & f(1) r_3 v_3 & f(2) r_3 v_4 \\
 4 & - & - & - & f(1) r_4 v_4
\end{array}
\end{displaymath}
\caption{\tbllabel{r estimate}Model of distribution of votes among
resolves in time intervals between successive resolve introductions,
here shown for the first four resolves.}
\end{table}

We estimate the $r_j$ and $f(a)$ values as those maximizing the
likelihood of getting the observed numbers of votes on the resolves in
these time intervals, coming from independent Poisson distributions.
This maximization does not have a simple closed form, but setting
derivatives with respect to these parameters to zero does give simple
relations between these values at the maximum:
\begin{eqnarray}
r_j &=& \frac{\vResolve_j}{\sum_{a=1}^{R-j+1} f(a) v_{a+j-1}}
\eqlabel{r} \\
f(a) &=& \frac{\vAge_a}{\sum_{j=1}^{R-a+1} r_j v_{a+j-1}} \eqlabel{f(a)}
\end{eqnarray}
where $\vResolve_j$ is the number of votes resolve $j$ has received,
$\vAge_a$ is the number of votes made to resolves of age $a$ at the
time of the vote, in both cases excluding the initial vote to each
resolve. The resulting $f(a)$ estimates from the numerical solution
are similar to the distribution of votes vs.~age in \fig{votes vs
age}, and \fig{r estimates} shows the distribution of estimated $r$
values and a lognormal fit.

\begin{figure}[t]
\centering \includegraphics[width=\figwidth]{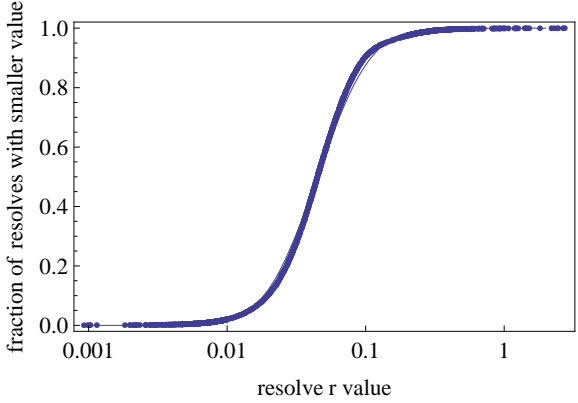}
\caption{\figlabel{r estimates}Cumulative distribution of $r$ values
for the resolves as obtained from a maximum likelihood estimate for
the observed data. The curve shows a lognormal distribution fit, with
parameters $\mu=-3.11\pm 0.01$ and $\sigma=0.69\pm 0.01$.}
\end{figure}

With the wide variation in $r$ values for resolves and the activity
rates for users (\fig{rho estimates}), a natural question is whether
these variations are related. In particular, whether the most active
users tend to preferentially introduce resolves that are especially
interesting to other users. While active users tend to introduce
more resolves overall, the correlation between the activity rate of
a user and the average $r$ values of the resolves introduced by that
user is small: $-0.06$. We find a modest correlation ($0.16$)
between the \emph{time} a user is active on the site and the mean
$r$ values of that user's introduced resolves.

To relate this model to the vote distribution of \fig{votes per
resolve}, consider the votes received by resolve $j$ up to and
including the time it is of age $A$. According to our model, the
number of votes, \emph{other than} its first vote, this resolve
receives is a Poisson variable $V_j(A)$ with mean
\begin{displaymath}
\mu_j(A) = r_j \sum_{a=1}^{A} f(a) v_{j+a-1}
\end{displaymath}
At the end of our data set, resolve $j$ is of age $R-j+1$.

The persistence of votes on resolves based on the wide variation of
$r$ values among resolves gives rise to a multiplicative process
with decay. To see this, in our model the number of votes between
successive resolve introductions is geometrically distributed with
mean $\mean{v} = (1 - q - \lambda) / q \approx 52$. Furthermore,
from \fig{votes vs age}, the aging function is approximately power
law, with $f(a) \approx a^{-s}$ and $\sum_{a=1}^{A} f(a) \sim
A^{1-s}/(1-s)$. The expected number of votes up to age $A$ is then
$\mu_j(A) \sim r_j \mean{v} A^{1-s}/(1-s)$. After accumulating many
votes (i.e., when $A$ is large), the actual number of votes $V_j(A)$
will usually be close to this expected value. The change in votes to
age $A+1$ is
\begin{eqnarray*}
V_j(A+1) & \approx & r_j \mean{v} \frac{A^{1-s}}{1-s}(1 + x) \cr
    & \approx & V_j(A) (1+x)
\end{eqnarray*}
where $x$ is a nonnegative random variable with mean $(1-s)/A$.
Thus, except possibly for the votes a resolve receives shortly after
its introduction, the growth in number of votes is well-described by
a multiplicative process with decay.

That our model corresponds to a multiplicative process has two
consequences. First, a sample obtained at a range of ages from a
multiplicative process (with or without decay) leads to the double
Pareto lognormal distribution seen in \fig{votes per resolve}. In
our case, the sample has a uniform range of ages from 1 to $R$,
though with the decay older resolves accumulate votes more slowly
than younger ones. A second consequence arises from the decay as
resolves become less visible over time. Thus our model provides one
mechanism using locally available information giving rise to
dynamics governed by multiplicative random variation with decay. A
similar process arises if the decay is due to any combination of
decreasing interest in the content and loss of visibility with age,
e.g., as seen in sites such as Digg~\cite{wu07} with current events
stories that become less relevant over time.

\begin{figure}[t]
\centering \includegraphics[width=\figwidth]{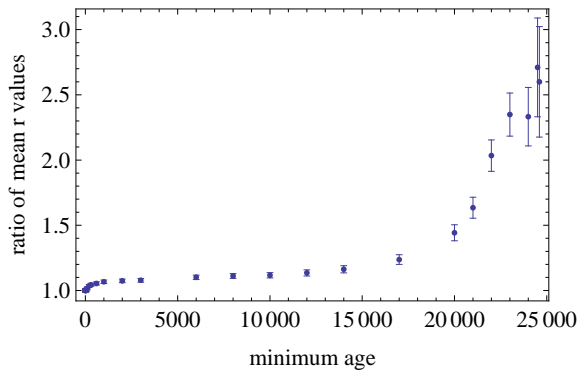}
\caption{\figlabel{avg r ratio by vote age}Ratio of means of $r$
estimates for resolves receiving votes at or after various ages to
the $r$ estimates for all resolves of those ages. Error bars
indicate the standard error in the ratio of means from the standard
deviation of the $r$ values and the number of resolves in each
category.}
\end{figure}

As one indication of the diversity of voting on resolves, \fig{avg r
ratio by vote age} shows how the average $r$ value for resolves
receiving votes compares to the average for all resolves among those
at least a given age. Randomization tests indicate average $r$
values of resolves receiving votes are unlikely to be the same as
those of all resolves at each of these ages, with $p$-values less
than $10^{-3}$ in all cases. With increasing age, resolves
continuing to receive votes tend to be those with especially high
$r$ values. This behavior indicates that high interestingness
estimates for resolves persist over time, as a small subset of
resolves continue to collect votes well after their introduction.

\section{Links}\sectlabel{links}

Users' decisions of who to link to and how they attend to the behavior
of their neighbors can significantly affect the performance of
participatory web sites. A common property of such networks is the
wide range in numbers of links made by users, i.e., the degree
distribution of the network. The structure of the networks is typical
of those seen in online social networking sites, and the links created
by users generally conform to their nominal semantics~\cite{hogg08a}.

The degree distributions in all three Essembly networks are close to
a truncated power law~\cite{newman01}, with number of users in the
network with degree $d$ proportional to $d^{-\tau} e^{-d/\kappa}$.
\fig{degreedistribs} shows the distribution of degrees in the
networks.

\begin{figure}[t]
\centering
\includegraphics[width=\figwidth]{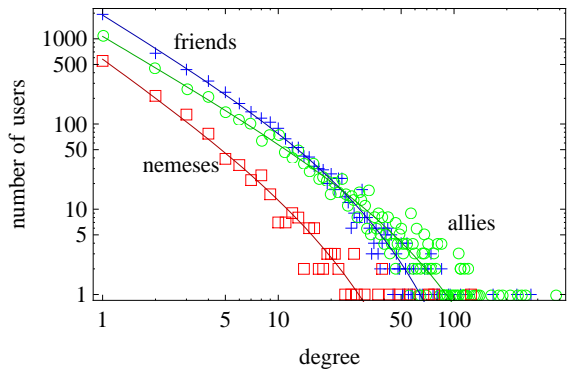}
\caption{\figlabel{degreedistribs}The number of users in the
Essembly
  social network who have a given number of links of the indicated
  type (plus symbols are for the friends, circles for the allies, and
  squares for the nemeses networks, respectively). The parameters of
  the best fits of truncated power laws on the three sets of data are
  given in the text.}
\end{figure}

These long-tail degree distributions are often viewed as due to a
preferential attachment process in which users tend to form links
with others in proportion to how many links they already have.
Combined with a limitation on the number of links a user has, this
process gives truncated power-law degree
distributions~\cite{amaral00}. For Essembly, this limitation arises
from users becoming inactive, since such users no longer accept
links. However, users in Essembly have no direct access to number of
links of other users. Thus we need to identify a mechanism users
could use, based on information available to them.

The mechanism underlying preferential attachment likely differs
between the friends network and the two ideological ones. In
particular, since the links appear to follow their nominal
semantics~\cite{hogg08a}, links in the friends network are likely to
be mainly between people who know each other (i.e., not found via
Essembly) while ideological links (especially those who are not also
friends) require finding the people by ideological profile (which
Essembly makes available). Building such a profile requires voting,
so a user with many votes is more likely to have voted on similar
resolves as other users. Such common votes allow ideological
comparisons between users and therefore suggestions for potential
users to link to.

The need to build ideological profiles suggests votes on common
resolves is key to the number and type of links. That is, a user
forming a link is more likely to have many common votes with other
users who are very active (and hence have many votes). Thus forming
links based on common votes is likely to lead people to link
preferentially to highly active users, who will in turn tend to have
many links.

One challenge to evaluating this mechanism is causation: resolves
voted on by network neighbors are highlighted in the user interface,
making them more visible and hence more likely to receive votes. Thus
common votes increase the chances of forming a link by providing
information to form a profile, and links increase the chance of common
votes through visibility of resolves. Separating these effects is
especially challenging since our data set does not indicate
\emph{when} each link was formed.

We can partially address this challenge through two observations.
First, Essembly presents ``resolves in your network'' grouping the
three networks together. So any influence on resolve visibility due
to networks should be similar for all networks. Second, \fig{common
votes} illustrates a distinction between the ideological networks in
Essembly and the social network nominally linking people who know
each other as friends. The figure shows friends generally have many
more resolves in common, i.e., both users voted on, than random
pairs of users who participate in at least one of the networks. The
figure also shows the ideological networks (both allies and nemeses)
are similar and have significantly more common resolves than the
friends network.

These two observations suggest the enhanced number of common votes
for the friends network compared to random pairs is primarily due to
the increased visibility of resolves due to network neighbors voting
on them. Because Essembly presents resolves from all networks
together, this enhancement is also likely to be the same for the
ideological networks. Hence, the remaining increase in common votes
in the ideological networks compared to the friends network suggests
the additional commonality required for users to form the links.

\begin{figure}[t]
\centering
\includegraphics[width=\figwidth]{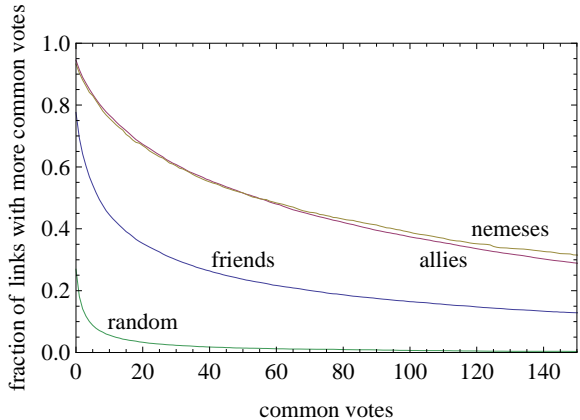}
\caption{\figlabel{common votes}Cumulative distribution of number of
common votes among linked pairs in the networks, and among random
pairs of users who are in at least one network. For each number of
common votes, the curves show the fraction of pairs with more than
that many resolves both users in the pair voted on.}
\end{figure}

\begin{figure}[t]
\centering
\includegraphics[width=\figwidth]{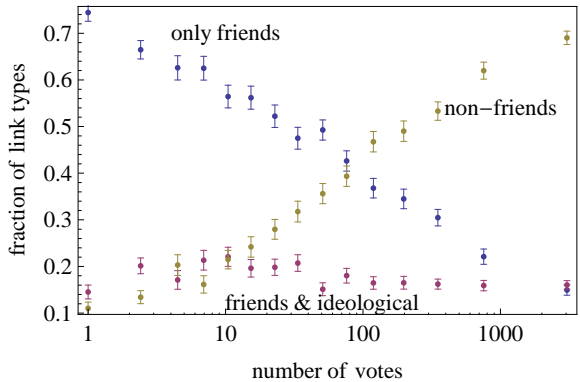}
\caption{\figlabel{link types}Fraction of link types vs.~number of
votes on noninitial resolves. A linked pair of users are denoted
``only friends'' when their only link is in the friends network,
``non-friends'' when they are not linked in the friends network, and
``friends \& ideological'' when they have a friends link as well as
a link in the allies or nemeses networks.}
\end{figure}

\fig{link types} shows the types of links vary depending on user
activity. For this plot, users with at least one network connection
are grouped into quantiles by their number of votes. Each point on
the plot is the average fraction of link types among users in that
quantile, with the error bar indicating the standard deviation of
this estimate of the mean. Users with few votes tend to have most of
their links to friends only, so do not participate much in the
ideological networks. On the other hand, users with many votes tend
to have most of their links in the ideological networks and to
people who are not also friends. The same trend in link types occurs
as a function of other measures of user activity, i.e., using
quantiles based on the time a user is active or the number of links
a user has.

\subsection{Model}

In our model, user $u$ forms links at a rate $\lambda \rho_u$. Thus
the number of links a user forms is a combination of activity rate
and how long the user remains at the site. The wide variation in
activity times and $\rho_u$ among users (\fig{activity time} and
\figbare{rho estimates}) gives rise to a wide distribution of number
of links.

While the most common mechanisms designed to reproduce the observed
power law degree distributions use growing rules and the degree of
vertices in link formation~\cite{dorogovtsev02}, in the following we
propose a mechanism that only takes into account the extent to which
two users share interests to describe link formation between two
users.

Because links involve two people, an additional modeling issue is
which pairs of users form links. In our model, we take the friends
network to primarily reflect a preexisting social network. For the
ideological networks, however, we take the choices to depend on
common votes. Furthermore, only active users can form links.
Specifically, we model the likelihood a (non-friend) pair forms a
link in an ideological network as proportional to the number of
common votes they have. In addition, existing friends links can add
an ideological link.

People form ideological links based on common votes, and only users
active at the same time can form links. The first factor gives more
links for those who vote a lot (due to being more likely to have
votes in common with others). This leads to, in effect, preferential
attachment for forming links (those with more links are likely to be
users with more votes, hence more overlap with others), while the
attachment probability does not explicitly depend on degrees. The
activity constraint limits the link growth, corresponding to
descriptive models giving truncated power-law degree
distribution~\cite{amaral00}.

\subsection{Behavior}

To verify whether users connect to each other based on similarities
in their voting profile, we propose the following simplified
mechanism for link formation. Suppose that user $A$ voted on $N_A$
resolves, while user $B$ voted on $N_B$ resolves in total. Assuming
that $A$ and $B$ form a link with a probability proportional to the
number of votes that the pair has in common, this probability will
be $P_l(A,B) \propto N_A N_B$ if $N_A$ and $N_B$ are sufficiently
smaller than the number of all resolves, and $A$ and $B$ vote
independently of each other and pick resolves randomly from the pool
of all available resolves. Caldarelli~et~al.~have shown that if
vertices in a network possess intrinsic ``fitnesses'', and the
linking probability is proportional to the product of fitnesses of
the two vertices to be linked, then in the particular case when the
fitnesses are drawn from a power-law probability distribution
function the resulting degree distribution will have the same
exponent as the fitness distribution~\cite{caldarelli02}. We can
consider the number of votes a person makes as the fitness of the
vertex, and arrive by analogy at the same model as
Ref.~\cite{caldarelli02}, resulting in an expected power-law
exponent of $-1.45$ (\fig{votes per user}).

Fitting truncated power laws to the degree distributions of the
three networks shown in \fig{degreedistribs}, we found the
parameters $\tau_F = 1.25 \pm 0.04$, $\kappa_F = 27 \pm 4$; $\tau_A
= 1.20 \pm 0.04$, $\kappa_A = 59 \pm 9$; and $\tau_N = 1.44 \pm
0.11$, $\kappa_N = 18 \pm 6$ for the friends, allies, and nemeses
networks, respectively, with the values for the $95\%$ confidence
intervals indicated. The power-law exponents are in the range
$[1.20, 1.44]$, giving a consistent match to the exponent of
\fig{votes per user}. The truncation of the power laws seen in the
degree distributions are most likely the result of vertices
gradually becoming inactive in time. An interesting consequence of
the above is that while the friends network as seen on Essembly is
supposed to not be a result of shared votes made conspicuous by the
web user interface, we see a consistent match in the exponents: this
suggests that friendship links in real life may also form around
shared interests, and that the scope of interests people have may
follow a similar probability distribution function as shown in
\fig{votes per user}.

Unlike random graph models with this degree
distribution~\cite{newman01}, our mechanism based on common votes
also gives significant transitivity, comparable to that observed for
the allies network. That is, if users $A$ and $B$ have voted on many
resolves in common, as have users $B$ and $C$, then users $A$ and
$C$ also tend to have significant overlap in the resolves they voted
on.

A further consequence of our model with ideological links depending
on common votes is the prediction of a change in the types of links
users make as they vote. In particular, users with few votes will
also have few common votes with other users and hence their links
will tend to be mostly friends. Users with many votes, on the other
hand, will tend to have common votes with many others and hence,
according to this model, tend to have mostly ideological links. This
change in type of links as a function of a user's number of votes or
links occurs in Essembly, as seen in \fig{link types}.

Finally, our model also describes the significant fraction of users
who form no links as due to a combination of low activity rate
$\rho$ and short activity time $t$. Specifically, in our model the
probability a user has no links is $e^{-\lambda \rho t}$. For active
users, whose activity time distribution is roughly exponential with
time constant $\tau$, the values in \tbl{user parameters} and the
distribution of $\rho$ values in \fig{rho estimates} give the
probability for no links as the average value of $e^{-\lambda \rho
t}$ equal to $23\%$. This compares with the 1242 out of 4762 active
users (i.e., $26\%$) who have no links in our data set.

\section{Online Estimation}\sectlabel{online estimation}

Our model allows estimating parameters for new users and new
resolves as they act in the system. In particular, we describe using
the early history of resolves to estimate the number of votes a
resolve will eventually have as well as which resolve will likely
receive the next vote.

\begin{figure}[t]
\centering
\includegraphics[width=2.5in]{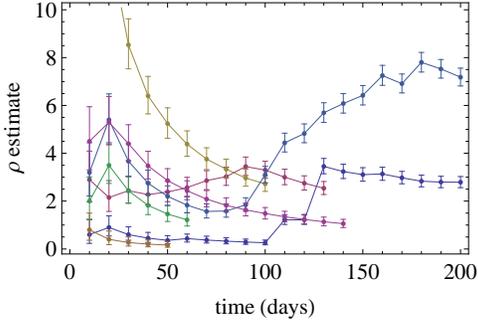}
\caption{\figlabel{rho estimates vs time}Estimates of $\rho$ values
for several users as a function of the time since their first vote.
Error bars show the $95\%$ confidence intervals.}
\end{figure}

\fig{rho estimates vs time} shows estimates of user activity levels as
a function of time since the user first voted. We see user activity
levels change with time, and in different ways. So users not only
differ considerably in their average activity rates but also in how
their interest in the site varies in time.

\begin{figure}[t]
\centering
\includegraphics[width=2.5in]{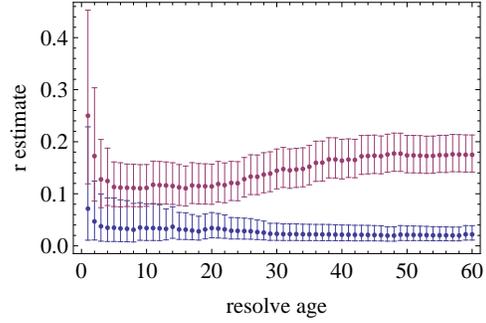}
\caption{\figlabel{r estimates vs age}Estimates of $r$ values for
two resolves as a function of their age. Error bars show the $95\%$
confidence intervals.}
\end{figure}

For resolves, using the model of \sect{resolves}, \fig{r estimates
vs age} shows how estimates for resolves, and their confidence
intervals change over time, as more votes are observed. Other
resolves show similar behavior. Thus the interestingness of resolves
appears to converge in time as we expect.

In practice, however, the optimization procedure is computationally
very costly due to the large number of parameters that grows
linearly with the number of resolves in the system. A further
requirement of an online algorithm is that it is able to update the
model parameters in real time as new users, votes and resolves enter
the system. Thus it is not feasible to consider a growing number of
resolves with constant resources. Instead we must limit the the
number of parameters and thus resolves to be optimized to a constant
value.

One such approach is to optimize parameters based on the last $K$
active resolves only, and keep the interestingness and aging
parameters constant for resolves older than that. This method,
interestingly, has the potential benefit of being able to track
changes to interestingness and aging in time.

Another incremental approach uses the observation that old resolves,
with a long track record of votes, have their interestingness
well-estimated and similarly the aging function $f(a)$ for small
ages is well-estimated from prior experience with many resolves
receiving votes at those ages. Conversely, recently introduced
resolves have had little time to accumulate votes and $f(a)$ for
large ages is poorly estimated due to having little experience in
the system with resolves that old. Furthermore, we can expect $f(a)$
to change slowly with time as primarily due to how the user
interface makes resolves visible to users. The maximum likelihood
estimation for these parameters described in \sect{resolves}
requires a computationally expensive optimization to find the best
choices for $r_j$ and $f(a)$ for all values. For new resolves, with
$j$ close to $R$, \eq{r} determines the $r_j$ values in terms of the
values of $f(a)$ for small ages (i.e., $a=1,\ldots,R-j+1$) which are
already well-determined from the prior history of the system. So
instead of an expensive reevaluation of all the $r$ and $f$ values,
we can simply incrementally estimate the $r$ values of new resolves
assuming $f(a)$ values for small ages do not change much.
Conversely, as new resolves are introduced, the oldest resolves in
the system advance to ever larger ages, allowing estimates of $f(a)$
for those ages from \eq{f(a)} by assuming the $r$ values of those
old resolves do not change much with the introduction of new
resolves.

Such estimates of model parameters can be useful guides for
improving social web sites if extended to user behavior as well, by
identifying new users likely to become highly active or content
likely to become popular. Since it is possible to estimate the
statistical errors given the sample size, one can also perform risk
assessment when giving the estimates. Newly posted content with high
interestingness, for instance, can be quickly identified and given
prominent attention on the online interface.

\section{Discussion}

We described several extended distributions resulting from user
behavior on Essembly, a web site where users create and rate content
as well as form networks. These distributions are common in
participatory web sites. From the extended distributions of user
behavior we find extremely heterogenous population of users and
resolves. We introduce a plausible mechanism describing user
behavior based on locally available information, involving a
combination of aging and a large variation among people and
resolves. We centered our investigations around three areas: the
wide range in user activity levels in online participation; how
online social networks form around topical interests; and the
factors that influence the popularity of user-created content.

In particular, we found, first, that most users try the online
services only briefly, so most of the activity arises from a
relatively small fraction of users who account for the diverse
behavior observed. Second, we gave a plausible, quantitative
explanation of the long-tailed degree distributions observed in
online communities, based on only the observed activity patterns of
users and the underlying collaborative mechanisms. Our observations
suggest different mechanisms underly the formation of the social
(friends) and ideological (allies and nemeses) networks, although
these mechanisms give similar outcomes, e.g., for the qualitative
form of the degree distribution. The implications may extend beyond
the scope of purely online societies to describe other societal
connections as well where shared interests motivate relationship
formation. Our model, however, does not address other significant
properties of the networks, such as community structure and
assortativity and why they differ among the three
networks~\cite{hogg08a}. Nor does our model address detailed effects
on user behavior due to their network neighbors. Third, we proposed
a model and algorithm that can describe and predict through
iterative refinements how the popularity of user-generated
submissions evolves in time, considering both their changing
exposure online and their inherent interestingness. We found that
the exposure that content receives depends largely on its recency,
and decays with age.

The characteristics of our models plausibly apply to other web sites
where user participation is self-directed and where content creation
and social link formation plays a dominant part in the individual
online activities. The Digg and Wikipedia user communities (those
whose activity data is publicly available) in particular may show
similar behavior in their activity patterns. Our models could be
extended to include the weak, but nevertheless statistically
significant, correlations among user behaviors such as activity rate
and the time they remain active on the site. Including such
correlations, as well as some historical and demographic information
on individual users, may improve the model predictions as seen, for
example, in models estimating customer purchase
activities~\cite{borle08}.

Consequences of our model include suggestions for identifying active
users and interesting resolves early in their history. E.g., from
persistence in voting rates over time, even before accumulating
enough votes to be rated as popular. Such identification could be
useful to promote interesting content on the web site more rapidly,
particularly in the case of niche interests. Beyond helping users
find interesting content, designs informed by causal models could
also help with derivative applications, such as collaborative
filtering or developing trust and reputations, by quickly focusing
on the most significant users or items. Such applications raise
significant questions of the relevant time scales. That is, observed
behavior is noisy, so there is a tradeoff between using a long time
to accumulate enough statistics to calibrate the model vs.~using a
short time to allow responsiveness faster than other proxies for
user interest such as popularity.

Our models raise additional questions on population properties we
used. One such question is understanding how the resolve aging
function relates to the user interface and changing interests among
the user population. Another question is how the wide distributions
in user activity and resolve interestingness arise. The lognormal
fits suggest underlying multiplicative processes are involved. It
would also be interesting to extend the model to identify niche
resolves, if any. That is, resolves of high interest to small
subgroups of users but not to the population as a whole.
Automatically identifying such subgroups could help people find
others with similar interests by supplementing comparisons based on
ideological profiles.

A caveat on our results, as with other observational studies of web
behavior, is the evidence for mechanisms is based on correlations in
observations. While mechanisms proposed here are plausible causal
explanations since they rely on information and actions available to
users, intervention experiments would give more confidence in
distinguishing correlation from causal relationships. Our model
provides testable hypotheses for such experiments. For example, if
intrinsic interest in resolves is a major factor in users' selection
of resolves, then deliberate changes in the number of votes may
change visibility but will not affect interestingness. In that case,
we would expect subsequent votes to return to the original trend.
Thus one area for experimentation is to determine how users value
content on various web sites. For example, if items are valued
mainly because others value them (e.g., fashion items) then observed
votes would \emph{cause} rather than just reflect high value. In
such cases, random initial variations in ratings would be amplified,
and show very different results if repeated or tried on separate
subgroups of the population. If items all have similar values and
difference mainly due to visibility, e.g., recency or popularity,
then we would expect votes due to rank order of votes (e.g., whether
item is most popular) rather than absolute number of votes. If items
have broad intrinsic value, then voting would show persistence over
time and similar outcomes for independent subgroups. It would also
be useful to identify aspects of the model that could be tested in
small groups, thereby allowing detailed and well-controlled
laboratory experiments comparing multiple interventions. Larger
scale experiments~\cite{bainbridge07,salganik06} would also be
useful to determine the generality of these mechanisms.

The key features of continual arrival of new users, existing users
becoming inactive and a wide range of activity levels among the user
population and interest in the content can apply in many contexts.
For the distribution of how user rate content (e.g., votes on
resolves in Essembly), generalizing to other situations will depend
on the origin of perceived value to the users. At one extreme, which
seems to apply to Essembly, the resolves themselves have a wide
range of appeal to the user population, leading some items to
consistently collect ratings at higher rates than others. At the
other extreme, perceived value could be largely driven by popularity
among the users, or subgroups of users, as seen in some cultural
products~\cite{salganik06}. In rapidly changing situations, e.g.,
current news events or blog posts, recency is important not only in
providing visibility through the system's user interface, but also
determining the level of interest. In other situations, the level of
interest in the items changes slowly, if at all, as appears to be
the case for resolves in Essembly concerning broad political
questions such as the benefits of free trade. All these situations
can lead to long-tail distributions through a combination of a
``rich get richer'' multiplicative process and decay with age. But
these situations have different underlying causal mechanisms and
hence different implications for how changes in the site will affect
user behavior. Thus, design and evaluation of participatory web
sites can benefit from the availability of causal models.

\small

\section*{Acknowledgments}
We thank Chris Chan and Jimmy Kittiyachavalit of Essembly for their
help in accessing the Essembly data. We have benefited from
discussions with Michael Brzozowski, Dennis Wilkinson, and Tam\'as
Sarl\'os.


\begin{thebibliography}{10}

\bibitem{agichtein06}
E.~Agichtein, E.~Brill, S.~Dumais, and R.~Ragno.
\newblock Learning user interaction models for predicting web search result
  preferences.
\newblock In {\em Proc. of the International ACM SIGIR Conference on Research
  and Development in Information Retrieval}, pages 3--10, 2006.

\bibitem{amaral00}
L.~A.~N. Amaral, A.~Scala, M.~Barthelemy, and H.~E. Stanley.
\newblock Classes of small-world networks.
\newblock {\em Proc. of the Natl. Acad. Sci.}, 97:11149--11152, 2000.

\bibitem{anderson06}
C.~Anderson.
\newblock {\em The Long Tail: Why the Future of Business is Selling Less of
  More}.
\newblock Hyperion, 2006.

\bibitem{bainbridge07}
W.~S. Bainbridge.
\newblock The scientific research potential of virtual worlds.
\newblock {\em Science}, 317:472--476, 2007.

\bibitem{boccaletti06}
S.~Boccaletti, V.~Latora, Y.~Moreno, M.~Chavez, and D.-U. Hwang.
\newblock Complex networks: Structure and dynamics.
\newblock {\em Physics Reports}, 424:157--308, 2006.

\bibitem{borle08}
S.~Borle, S.~S. Singh, and D.~C. Jain.
\newblock Customer lifetime value measurement.
\newblock {\em Management Science}, 54:100--112, 2008.

\bibitem{caldarelli02}
G.~Caldarelli, A.~Capocci, P.~{De Los Rios}, and M.~A. Munoz.
\newblock Scale-free networks from varying vertex intrinsic fitness.
\newblock {\em Physical Review Letters}, 89:258702, 2002.

\bibitem{carterette07}
B.~Carterette and R.~Jones.
\newblock Evaluating search engines by modeling the relationship between
  relevance and clicks.
\newblock In J.~Platt et~al., editors, {\em Advances in Neural Information
  Processing Systems}. NIPS, 2007.

\bibitem{clarke07}
C.~L.~A. Clarke, E.~Agichtein, S.~Dumais, and R.~W. White.
\newblock The influence of caption features on clickthrough patterns in web
  search.
\newblock In {\em Proc. of the International ACM SIGIR Conference on Research
  and Development in Information Retrieval}, pages 135--142, 2007.

\bibitem{craswell08}
N.~Craswell, O.~Zoeter, M.~Taylor, and B.~Ramsey.
\newblock An experimental comparison of click position-bias models.
\newblock In {\em Proc. of the International Conference on Web Search and Web
  Data Mining}, pages 87--94, NY, 2008. ACM.

\bibitem{dorogovtsev02}
S.~N. Dorogovtsev and J.~F.~F. Mendes.
\newblock Evolution of networks.
\newblock {\em Advances In Physics}, 51:1079--1187, 2002.

\bibitem{fortunato06}
S.~Fortunato, A.~Flammini, F.~Menczer, and A.~Vespignani.
\newblock Topical interests and the mitigation of search engine bias.
\newblock {\em Proc. of the Natl. Acad. of Sciences}, 103:12684--12689, 2006.

\bibitem{guha04}
R.~Guha, R.~Kumar, P.~Raghavan, and A.~Tomkins.
\newblock Propagation of trust and distrust.
\newblock In {\em Proc. of the 13th Intl. World Wide Web Conf. (WWW2004)},
  pages 403--412, New York, 2004. ACM.

\bibitem{hogg08a}
T.~Hogg, D.~M. Wilkinson, G.~Szabo, and M.~Brzozowski.
\newblock Multiple relationship types in online communities and social
  networks.
\newblock In {\em Proc. of the AAAI Symposium on Social Information
  Processing}, 2008.

\bibitem{huberman99a}
B.~A. Huberman and L.~A. Adamic.
\newblock Growth dynamics of the {World Wide Web}.
\newblock {\em Nature}, 401:131, 1999.

\bibitem{huberman98}
B.~A. Huberman, P.~L.~T. Pirolli, J.~E. Pitkow, and R.~M. Lukose.
\newblock Strong regularities in {World Wide Web} surfing.
\newblock {\em Science}, 280:95--97, 1998.

\bibitem{james07}
A.~James and M.~J. Plank.
\newblock On fitting power laws to ecological data.
\newblock arxiv.org preprint 0712.0613, 2007.

\bibitem{joachims05}
T.~Joachims, L.~Granka, B.~Pan, H.~Hembrooke, and G.~Gay.
\newblock Accurately interpreting clickthrough data as implicit feedback.
\newblock In {\em Proc. of the International ACM SIGIR Conference on Research
  and Development in Information Retrieval}, pages 154--161, 2005.

\bibitem{lam04}
C.~Lam.
\newblock {SNACK:} incorporating social network information in automated
  collaborative filtering.
\newblock In {\em Proc. of the 5th ACM Conference on Electronic Commerce
  (EC'04)}, pages 254--255. ACM Press, 2004.

\bibitem{lempel03}
R.~Lempel and S.~Moran.
\newblock Predictive caching and prefetching of query results in search
  engines.
\newblock In {\em Proc. of the International Conference on World Wide Web},
  pages 19--28, 2003.

\bibitem{lerman07}
K.~Lerman.
\newblock Social information processing in social news aggregation.
\newblock arxiv.org preprint cs.cy/0703087, 2007.

\bibitem{lerman07a}
K.~Lerman.
\newblock User participation in social media: {Digg} study.
\newblock In {\em IEEE/WIC/ACM Intl. Conf. on Web Intelligence and Intelligent
  Agent Technology}, pages 255--258, 2007.

\bibitem{mitzenmacher04}
M.~Mitzenmacher.
\newblock A brief history of generative models for power law and lognormal
  distributions.
\newblock {\em Internet Mathematics}, 1:226--251, 2004.

\bibitem{newman05}
M.~E.~J. Newman.
\newblock Power laws, {Pareto} distributions and {Zipf's} law.
\newblock {\em Contemporary Physics}, 46:323--351, 2005.

\bibitem{newman01}
M.~E.~J. Newman, S.~H. Strogatz, and D.~J. Watts.
\newblock Random graphs with arbitrary degree distributions and their
  applications.
\newblock {\em Physical Review E}, 64:026118, 2001.

\bibitem{reed04}
W.~J. Reed and M.~Jorgensen.
\newblock The double {Pareto}-lognormal distribution: A new parametric model
  for size distributions.
\newblock {\em Communications in Statistics: Theory and Methods},
  33:1733--1753, 2004.

\bibitem{salganik06}
M.~J. Salganik, P.~S. Dodds, and D.~J. Watts.
\newblock Experimental study of inequality and unpredictability in an
  artificial cultural market.
\newblock {\em Science}, 311:854--856, 2006.

\bibitem{shockley57}
W.~Shockley.
\newblock On the statistics of individual variations of productivity in
  research laboratories.
\newblock {\em Proc. of the IRE}, 45:279--290, 1957.

\bibitem{vazquez03}
A.~V\'azquez.
\newblock Growing network with local rules: Preferential attachment, clustering
  hierarchy, and degree correlations.
\newblock {\em Physical Review E}, 67:056104, 2003.

\bibitem{vazquez06}
A.~V\'azquez, J.~G. Oliveira, Z.~Dezso, K.-I. Goh, I.~Kondor, and
A.-L.
  Barabasi.
\newblock Modeling bursts and heavy tails in human dynamics.
\newblock {\em Physical Review E}, 73:036127, 2006.

\bibitem{wu07}
F.~Wu and B.~A. Huberman.
\newblock Novelty and collective attention.
\newblock {\em Proc. of the Natl. Acad. Sci.}, 104:17599--17601, 2007.

\end{thebibliography}

\end{document}